# An Arcminute Resolution Search at 4.7 cm$^{-1}$ for Point Sources in MSAM and MAX Fields

S.E. Church[1], P.D. Mauskopf[1], P.A.R. Ade[2], M.J. Devlin[1],

W.L. Holzapfel[1], T.M. Wilbanks[1], A.E. Lange[3]

astro-ph/9409020   9 Sep 1994

[1]Department of Physics, University of California, Berkeley, CA 94720

[2]Department of Physics, Queen Mary and Westfield College, Mile End Road, London, E1 4NS, U.K.

[3]Department of Physics, California Institute of Technology, Pasadena, CA 91125

## ABSTRACT

We have searched for mm-wave emission from compact objects in two fields, each approximately 1 square degree in size, taken from regions of the sky in which degree-scale structure in the Cosmic Background Radiation (CBR) has recently been reported. The observations were made at frequency of 4.7 cm$^{-1}$ and with an angular resolution of 1$'$.7 using the Sunyaev-Zel'dovich Infrared Experiment (SuZIE) bolometer array at the Caltech Submillimeter Observatory (CSO). The first field was centered on 14$^{\rm h}$.92 +82° (1994.0), one of two regions in which Cheng et al. (1994) identify the signature of an unresolved point source seen during 0.5 degree resolution observations made at 5.6 cm$^{-1}$ with the Medium Scale Anisotropy Measurement (MSAM) experiment. The second field was centered on 15$^{\rm h}$.47 +72.4 (1994.0), part of the Gamma Ursae Minoris (GUM) region where several prominent features have been detected by Devlin et al. (1994) in 0.55 degree-resolution observations made at 3.5 and 6 cm$^{-1}$ with the Millimeter Anisotropy eXperiment (MAX). We find that there is no point source in either field that can account for the structure observed at 0.5 degree resolution, and that the structure must arise from objects with an angular size greater than 2$'$.

*Subject headings:* cosmic microwave background – cosmology: observations

## 1.   Introduction

Two experiments have recently reported structure in the Cosmic Background Radiation (CBR) on angular scales of less than a degree. The MAX experiment (Fischer et al. 1992) has reported structure with an amplitude of $\Delta T/T = 3.6 \pm 0.4 \times 10^{-5}$ (including systematic and statistical errors) for a gaussian autocorrelation function (GACF) with a $25'$ coherence angle in five observations distributed among three regions of sky (Clapp et al. 1994). The MSAM experiment has detected structure attributed to CBR anisotropy in an observation of a long strip at dec=82°. Cheng et al. identify the signature of an unresolved source in two regions of the strip and delete these points from the analysis. They argue that the amplitudes of these sources are inconsistent with CBR fluctuations obeying gaussian statistics and that they may be the result of emission from foreground sources. After these regions are deleted, the amplitude of the structure at $5.6$ cm$^{-1}$ is $\Delta T/T = 1.4 \pm 0.4 \times 10^{-5}$ for a GACF with an $18'$ coherence angle, significantly lower than the value measured by MAX. If both regions are included in the analysis, the amplitude is $\Delta T/T = 3.1 \pm 0.6 \times 10^{-5}$ (Cheng et al, 1994), in better agreement with the MAX results.

We have searched for compact (angular extent less than $2'$) sources of emission at $4.7$ cm$^{-1}$ in a one square degree field centered on the region $14^{\text{h}}\!.92 + 82$ (1994.0) in which Cheng et al. (1994) report the detection of the brighter of the two point sources, with a flux of $4.5 \pm 0.7$ Jy at $5.6$ cm$^{-1}$. We have also surveyed a similar size field centered on $15^{\text{h}}\!.47 + 72.4°$ (1994.0), one of several fields in the GUM region in which Devlin et al. (1994) observe prominent features.

## 2.   The Instrument

The observations were made using the Sunyaev-Zel'dovich Infrared Experiment (SuZIE) bolometer array at the Caltech Submillimeter Observatory (CSO) on Mauna Kea. The SuZIE array is designed to measure the Sunyaev-Zel'dovich (SZ) effect in clusters of galaxies and was recently used to make the first significant detection of the effect at millimeter wavelengths (Wilbanks et al., 1994a). The instrument, described in detail elsewhere (Wilbanks et al. 1994b), comprises a $2 \times 3$ array of bolometers operated at 300mK. Because of its relatively large beamsize (of order 2′) and high sensitivity ($< 200$ mJy/Hz$^{-1/2}$ at 4.7 cm$^{-1}$), this instrument is ideally suited for fast mapping of areas of sky covering a square-degree or so to a sensitivity of better than 1 Jy. A schematic of the array as it views the sky is shown in Figure 1. Removal of atmospheric and telescope emission is carried out by differencing each detector in a row against each of the other detectors in the same row. A novel form of electronic differencing is used in which the two bolometers are placed in an AC bridge, the output of which is synchronously demodulated to produce a stable DC signal corresponding to the brightness difference on the sky (Wilbanks et al. 1990). In terms of atmospheric subtraction, this is equivalent to a square-wave chop on the sky at infinite frequency. Two differences of 2′.2 and one of 4′.3 are obtained for each row of the array.

## 3. The Observations

The SuZIE array was used at 4.7 cm$^{-1}$ in April 1994 to make measurements of the SZ effect in several clusters of galaxies (Holzapfel et al. 1994). The surveys of the MSAM and GUM regions were carried out during these observations.

To make the observations, the array was oriented as shown in Figure 1, with the long axis parallel to the horizon. The CSO was then scanned in azimuth over the region of interest at a rate of 1 arcmin/sec, and the output from each AC bridge sampled every 0.2 s.

At this rate of scanning, a point source takes roughly 2 s to pass through each beam of the array, a time that is much greater than the 0.1-0.2 s time constant of the detector. Scans were made over a fixed range in azimuth of $\pm 30'$ and successive scans were separated by $1'$ in elevation. Observations of a complete *field* (either MSAM 15+82 or GUM 15.5+72.4) were broken into three *regions* each roughly $60' \times 22'$, but skewed by sky rotation. Mapping a complete field took about 1.5 hr, after which calibration observations of Uranus were made using the standard SuZIE observing mode. In this mode, the telescope is parked ahead of the source which then drifts through the field of view of the array. This method was used to obtain a high signal/noise map of the beams in azimuth (Uranus is $3\overset{''}{.}5$ in diameter and thus is unresolved by the SuZIE array). Additionally these observations provided absolute calibration of the measured flux; a 4.7 cm$^{-1}$ flux of 16 Jy for Uranus, based on the measurements of Griffin and Orton (1993), was assumed. The zenith opacity was estimated as 0.03 during these observations by using measurements from the CSO 225 GHz $\tau$ monitor and scaling to obtain $\tau$ values appropriate for 2.1 mm.

## 4. Data analysis

As the rms noise in a single scan is less than 0.5 Jy, a point source of 3 Jy or more would have been very noticeable in the raw data. An initial search by eye revealed no such strong sources in the raw scans. Spikes caused by cosmic ray hits were then removed from the data using an algorithm that carrys out a point-by-point differentiation on a scan and then looks for the large changes in slope that are characteristic of spike edges. About 5% of the data are removed by this algorithm. Tests showed that the signal from a point source with a flux less than 3 Jy would not be affected by this process.

Both of the regions observed are at high declination and do not reach very high elevations when observed at Mauna Kea (27° for the MSAM source and 30-35° for the GUM source). Consequently, the sensitivity of the survey is strongly limited by the extent to which atmospheric fluctuations can be removed from the data. The AC bridge technique performs a highly effective first differencing of the data, but it was found that adding a further level of differencing by combining pairs from the two rows significantly improved the rms noise.

Denoting the signal from each pixel on the sky as $c_i$, the output from each bridge can be written as $d_{i,j} = c_i - c_j$ with the differences $d_{1,2}$, $d_{2,3}$, $d_{3,1}$, $d_{4,5}$, $d_{5,6}$ and $d_{6,4}$ being sampled during each scan. Combining $d_{3,1}$ and $d_{6,4}$ in such a way as to generate a quadrupole beam on the sky was found to remove most of the residual atmospheric fluctuations in the scans. During these observations, $c_5$ was found to be suffering from excess noise which caused poor signal/noise in $d_{4,5}$ and $d_{5,6}$. Consequently the quadrupole difference could only be usefully formed from $d_{3,1}$ and $d_{6,4}$. In any case, the differences corresponding to the $4\overset{.}{.}3$ chop contain information on the widest range of angular scales and so are the most useful for a survey of this kind. To gain maximum benefit from the quadrupole chop procedure, the quadrupole difference, $q$, corresponding to the $i$th point in the scan is calculated using the form:

$$q(i) = d_{3,1}(i) + g \times d_{6,4}(i + \Delta)$$

where $g$ is a gain factor and $\Delta$ is a position offset in azimuth. Both $g$ and $\Delta$ were fixed for the analysis of a single field and the optimum values for each of the two fields were determined by minimizing the rms of $q$ calculated over the entire field. The offset, $\Delta$, arises from the interaction of the wind speed and direction with the scanning speed and direction of the array. For the MSAM scan, $\Delta = -4$ and for the GUM scan, $\Delta = 0$, were found to be the optimum values. The gain factor $g$ reflects variations in gain values across the array and also variations in the common mode rejection ratio between the two differences. The best values for the MSAM and GUM scans were found to be 0.74 and 0.85 respectively.

If a point source is present in either field then it would have been seen first by row

456, then two scans later by row 123. Of course, rotation of the source relative to the scan orientation will occur between these two observations; the effect is largest at the edges of the scans and even here is quite small. A point source at a relative azimuth of $\pm 30°$ that passes directly through the center of each pixel as row 456 is scanned across it, will have an amplitude that is reduced by 12% by sky rotation when observed by row 123 two scans later.

In order to search for point sources within the data, a model template is generated for each differenced pair by convolving a source that is very much smaller than the beam width with the beam profiles determined by drift scans over Uranus. The templates for $d_{3,1}$ and $d_{6,4}$ are then differenced in exactly the same way as the data to produce a model that can be fitted to the quadrupole-differenced data. In order to make use of the powerful constraint that any source must appear in two scans separated by $2'$ in elevation, two model templates are generated, one for a source observed by row 456 and the other for a source observed by row 123. A simultaneous fit of the first template to scan $n$ and the second template to scan $n + 2$ can then be carried out. The amplitude of the source in the data is thus determined by minimizing $\chi^2$ for each coupled pair of scans where:

$$\chi^2 = \sum_{i=1}^{N}(y_{i,n} - Am_{i,n} - B_n - C_n t_{i,n})^2 + \sum_{i=1}^{N}(y_{i,n+2} - Am_{i,n+2} - B_{n+2} - C_{n+2}t_{i,n+2})^2$$

Here $y_{i,n}$ is data from the $n$th scan in Volts, $A$ is the source flux in Jy, $m_{i,n}$ is the source template for scan $n$ in V/Jy, $B_n$ is a DC offset in Volts, $C_n$ is a time-dependent gradient in the data in V/sec and $t_{i,n}$ is the time in the scan in sec. This fit is then repeated using a template generated for a source centered at every position in scan $n$ in turn, yielding the best fit amplitude versus source position in the scan. In order to check this procedure, a 4.5 Jy test source was inserted into the MSAM data and then picked out by the fitting procedure with a signal/noise of 28. Thus, the MSAM source cannot be a point source with a spectrum that is flat between 5.6 and 4.7 cm$^{-1}$ since it would then have been clearly visible in our data (Figure 2 shows a section from an MSAM scan with the calculated response of the instrument to a 4.5 Jy source overlaid). A similar strength flat-spectrum point source would also be necessary to cause the structure seen in the GUM data. Again, no such source is visible in the raw scans.

We next considered the possibility that the observed structure may be due to a point source with a positive spectral index, sufficient to prevent it from being easily visible in our raw data. To determine whether faint point sources were present in any of the fields observed, the rms of the derived point source amplitudes was determined for each region and any amplitude that exceeded $3\sigma$ was tagged as the position of a potential source. Within each region, between 2 and 6 such 'sources' were detected, distributed equally as positive and negative amplitudes, suggesting that these are simply the tail of the distribution of fitted amplitudes. No source was detected in the MSAM error box above $3.5\sigma$ and no source was detected in the GUM error box above $4\sigma$. Within each region there are roughly 700 independent beam-sized patches. If the noise were gaussian-distributed then one would expect 1.8 points per region above the $3\sigma$ level. The region with the largest number of 'detections' (6 in MSAM region C) is also the region with the highest level of residual noise, suggesting that these detections are the result of a non-gaussian tail to the distribution of fitted fluxes, arising from atmospheric emission variations. In order to include these uncertainties in the data, we choose to express the results by giving a $3\sigma$ limit for each region, listed in Table 1, but we also list the maximum fitted source amplitude in the entire field. Figure 3 shows the regions of sky covered by these limits. For the MSAM field, the 30' square box shown in the figure is the error box quoted by Cheng et al. (1994) for the position of the unresolved source in their data; our coverage of this error box is about 99%. The 33' error box shown over the GUM regions reflects the size of the MAX beam. For a point source to be able to both produce the MSAM structure at 5.6 cm$^{-1}$ and to have a 4.7 cm$^{-1}$ flux of less than 1 Jy implies a spectral index, $\alpha > 8.6$ (at greater than 99% confidence). Clearly this is physically unrealistic. Since the 4.7 cm$^{-1}$ flux limits are significantly better in the GUM region, a point source with a positive spectral index is also unlikely as the source of the structure observed by MAX. To summarize, we see no evidence for a point source of the brightness that would be necessary to explain the MSAM unresolved source, or that would be sufficient to significantly contaminate the MAX observations of the GUM field.

Though we can rule out the hypothesis that the structure observed by MSAM or MAX in the two fields we have surveyed is due to a single source that is unresolved at 1.'7, it remains possible that the structure could be due to several compact sub-Jy sources clustered in the field, or that it may arise from a single source several arcminutes in extent. We have addressed the latter possibility by repeating the fitting procedure using different sized gaussian sources convolved with the SuZIE beam. Figure 4 shows the $3\sigma$ limit as a function of source full-width half-maximum for the regions observed. The $1\sigma$ limits on the flux of MSAM 15+82 from the MSAM measurement are also shown. If the source is assumed to have a spectrum that is flat between 4.7 and 5.6 cm$^{-1}$, then this plot can be used to infer a lower limit of 2.'8 to the size of such a source.

## 5.   Conclusions

We have surveyed two regions of sky covering an area of approximately one square degree each and centered on fields observed by the MSAM and MAX experiments. We have set an upper limit on the 4.7 cm$^{-1}$ flux from any point source in these fields of $S < 1$ Jy in MSAM 15+82 and $S < 0.6$ Jy in GUM 15.5+72.4. In particular we have been able to rule out a single point source as the origin of the unresolved feature MSAM 15+82 since, using any realistic model for the spectral index, such a source should have a flux of several Jy when observed with the SuZIE array. By carrying out fits of larger sources to our data, we conclude that if the feature has a flat spectrum at these wavelengths, then it must be larger than 2.8 in order to be consistent with both the MSAM and SuZIE observations.

Future experiments designed to measure intermediate-scale CBR anisotropy will have sensitivity of $\Delta T/T$ of $10^{-6}$ (Lange et al. 1994). Compact sources with mm-wave fluxes of 0.1 Jy or less could be a significant source of confusion for these experiments. Though all known sources of foreground confusion can, in principle, be distinguished from CBR anisotropy by their mm-spectra, the spectra of compact mm-wave sources are not well understood, and could mimic CBR anisotropy to the precision of the measurements. The observations that we report here illustrate the power of using a large aperture telescope to survey target fields for emission from compact sources. A future array, SuZIE II, will add multi-frequency coverage and increase several-fold the speed with which such regions can be surveyed.

This work has been made possible by a grant from the David and Lucile Packard Foundation, and by a National Science Foundation Presidential Young Investigator award to AEL. It has benefitted from useful discussions with Ken Ganga. We thank Jocelyn Keene and Gene Serabyn for providing the tertiary optics used at the CSO, and Anthony Schinckel and the entire staff of the CSO for their excellent support during the observations. The CSO is operated by the California Institute of Technology under funding from the National Science Foundation, Contract #AST-93-13929.

|                                   | MSAM 15+82 | GUM 15.5+72.4 |
|-----------------------------------|------------|---------------|
| Region A                          | 0.80 Jy    | 0.41 Jy       |
| Region B                          | 0.51 Jy    | 0.38 Jy       |
| Region C                          | 0.89 Jy    | 0.48 Jy       |
| Absolute point source flux limit  | $S < 1$ Jy | $S < 0.6$ Jy  |

Table 1: $3\sigma$ limits on point sources in the MSAM and GUM regions.

---

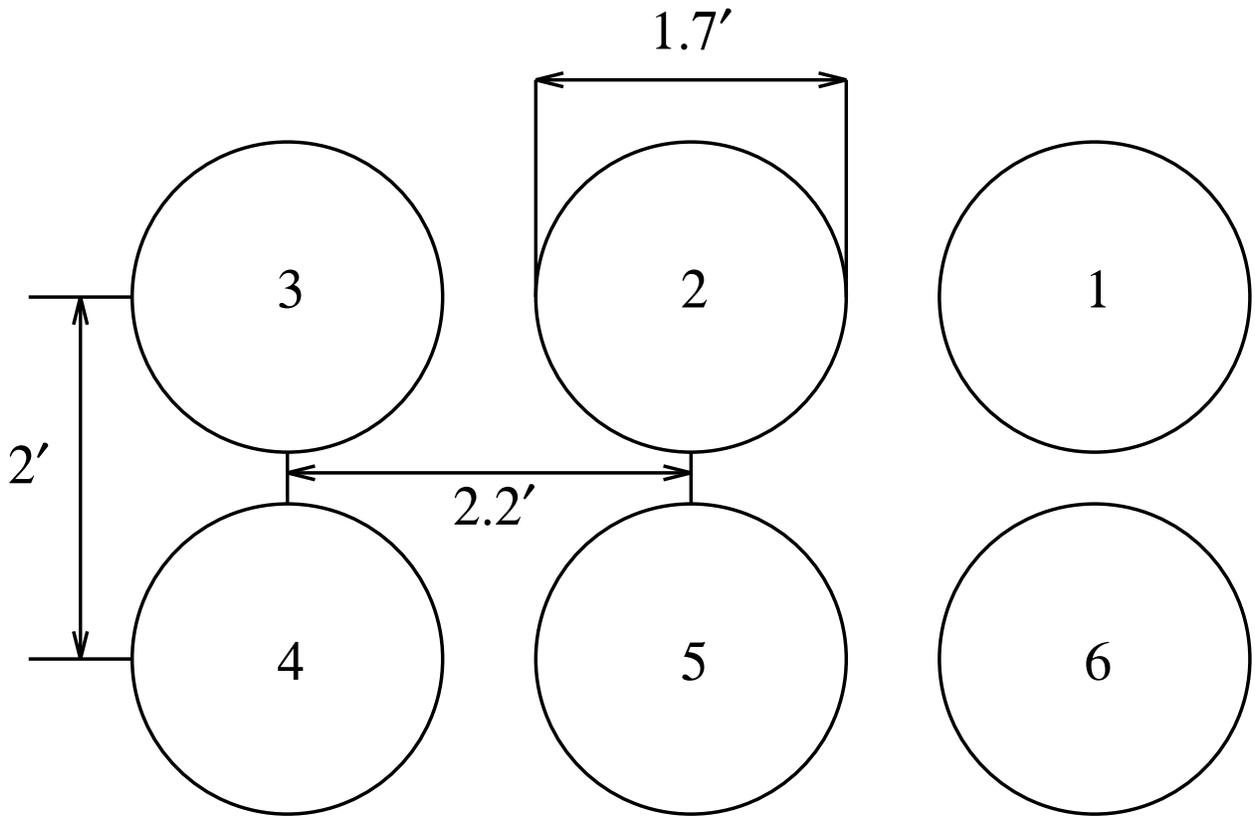

Fig. 1.— Schematic of the SuZIE array as it views the sky. During observations of MSAM and GUM regions, the long axis of the array was oriented parallel to the horizon with row 123 observing to the north of row 456

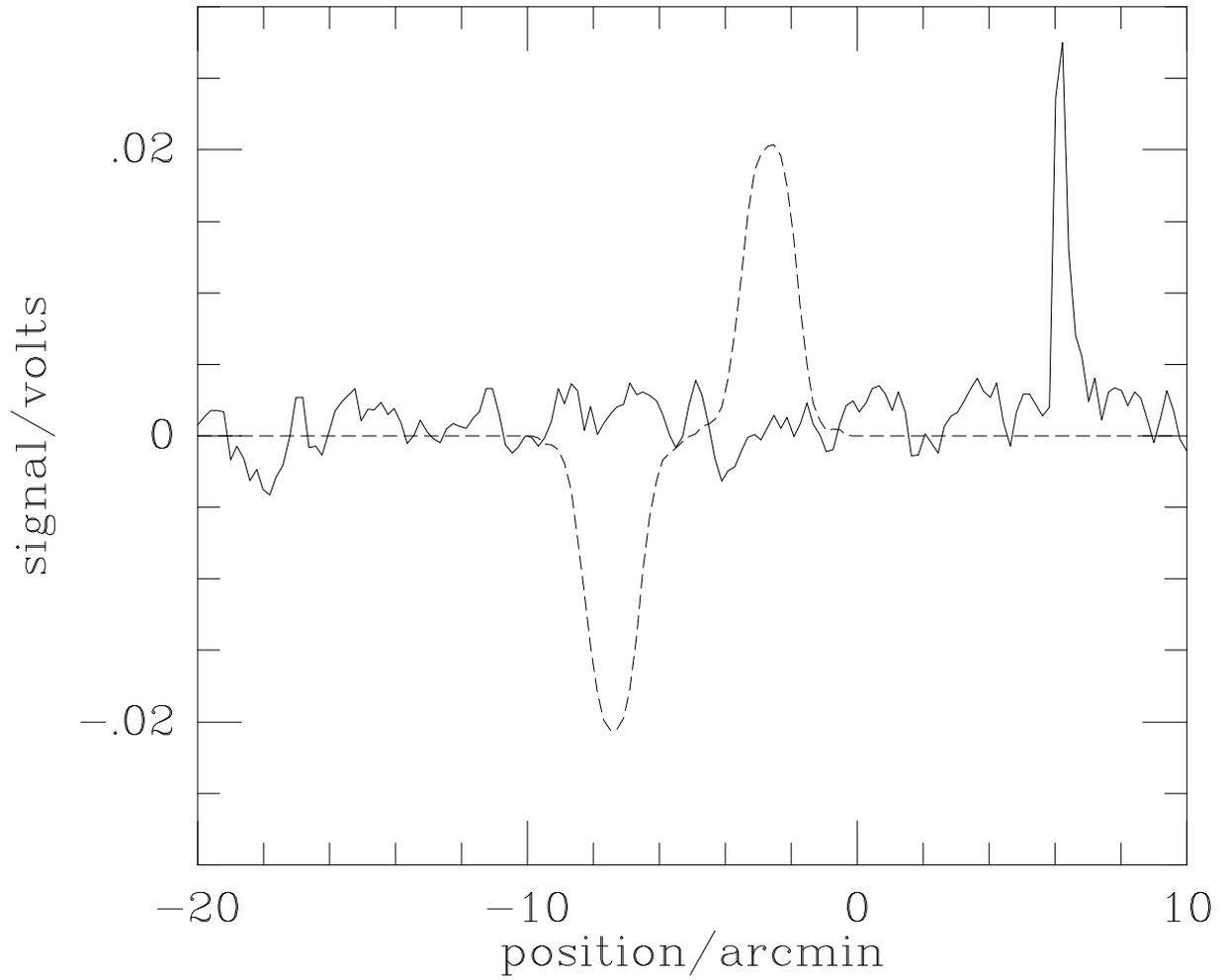

Fig. 2.— A portion of a single MSAM quadrupole-differenced scan with the calculated instrument response to a 4.5 Jy point source overlaid for comparison. The spike at a position offset of 6′ is the result of a cosmic ray hit.

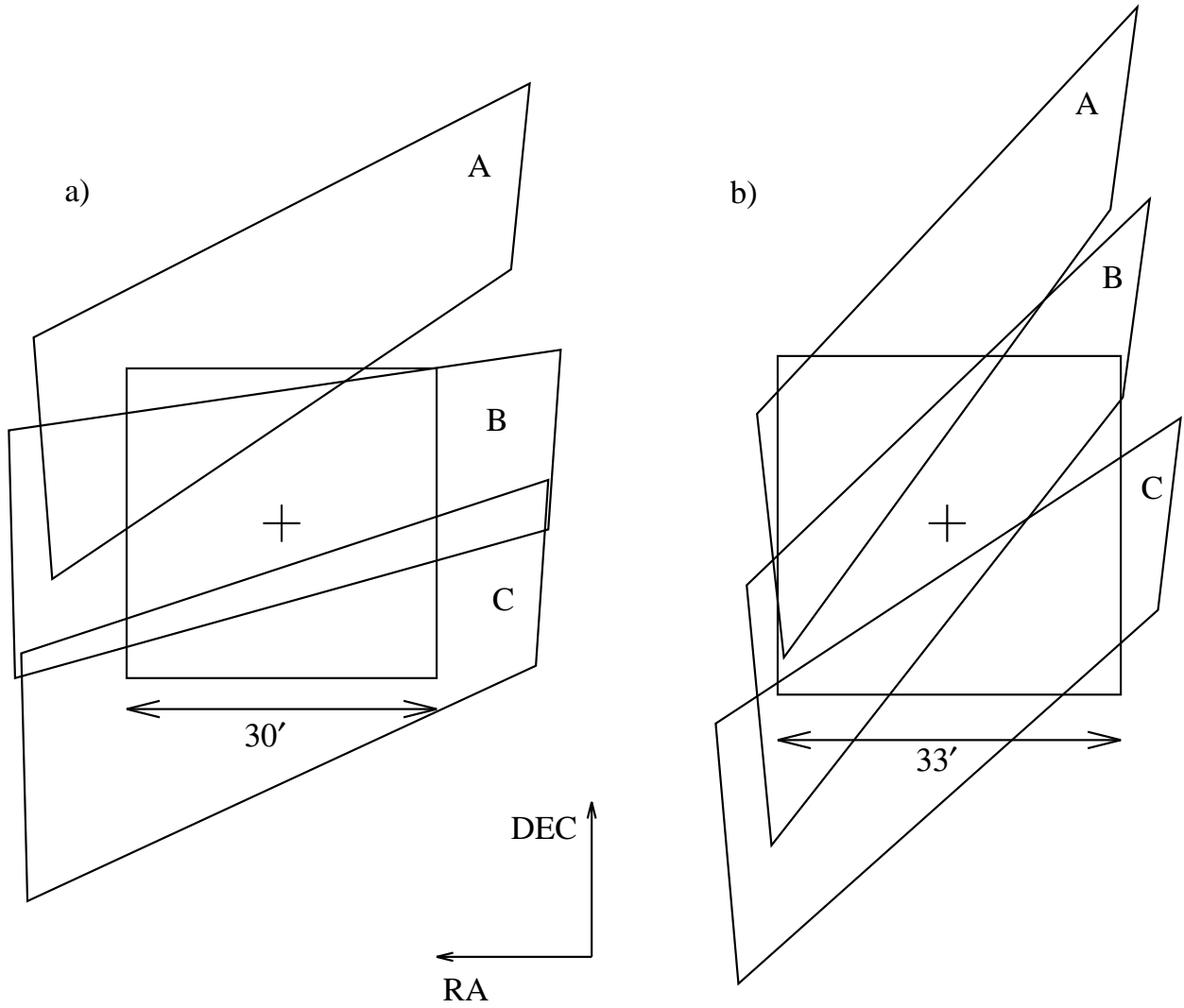

Fig. 3.— Schematic of a) the scanned regions centered on 15+82 with the quoted MSAM error box overlaid b) the scanned regions centered on 15.5+72.4 with an error box with the dimensions of the MAX beam overlaid.

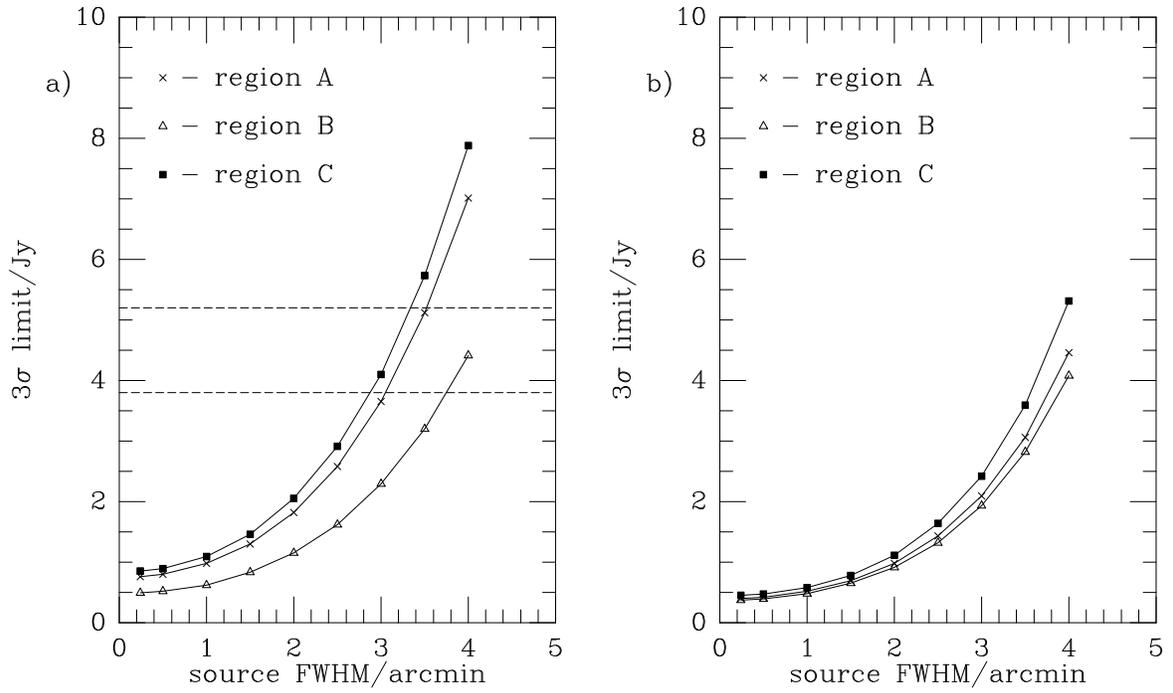

Fig. 4.— $3\sigma$ upper limit as a function of the full-width half-maximum of a fitted source with a gaussian profile for a) the MSAM and b) the GUM field.